\documentstyle[aps,eqsecnum,epsfig]{revtex}
\begin{document}
\draft
\title{Polyelectrolyte chains in poor solvent. A variational
  description of necklace formation.}
\author{Gabriele Migliorini $^{1}$ Vakhtang Rostiashvili$^{1}$  Namkyung Lee
  $^{1}$ and Thomas A. Vilgis$^{1,2}$}

\address{$^1$ Max Planck Institute for Polymer Research\\
  10 Ackermannweg, 55128 Mainz, Germany.\\
  $^2$ Laboratoire Europ\'een Associ\'e, Institute Charles Sadron\\
  6 rue Boussingault, 67083 Strasbourg Cedex, France.}
\date{\today}
\maketitle
\begin{abstract}
  We study the properties of polyelectrolyte chains under
  different solvent conditions, using a variational technique.  The free
  energy and the conformational properties of a polyelectrolyte chain are
  studied minimizing the free energy $F_N$, depending on $N(N-1)/2$ trial
  probabilities that characterize the conformation of the chain. The Gaussian
  approximation is considered for a ring of length $2^4<N<2^{16}$ and for an
  open chain of length $2^4<N<2^9$ in poor and theta solvent conditions,
  including a Coulomb repulsion between the monomers. In theta solvent 
  conditions the blob size is measured and found in agreement with
  scaling theory, including charge
  depletion effects, expected for the case of an open chain. In poor solvent
  conditions, a globule instability, driven by electrostatic repulsion, is
  observed.  We notice also inhomogeneous behavior of the monomer--monomer
  correlation function, reminiscence of necklace formation in poor
  solvent polyelectrolyte solutions. A global phase diagram in terms of solvent
  quality and inverse Bjerrum length is presented.
\end{abstract}

\section*{Introduction}

Polyelectrolytes are macromolecules with ionizable groups that can dissociate
in solution to produce polyions (polymers with fixed or mobile charges) that
are either positively or negatively charged and counter-ions of opposite
charge in solution. Polyelectrolytes are found in many biological systems
(proteins, nucleic acids, ..)  as well as in synthetic products (resins, ..).
As an example of polyelectrolyte we mention the polystyrene sulfonate $(C_8
H_7 SO_3 Na)_n$ that dissociates in water to produce polyanions $(C_8 H_7
SO_3^{-})_n$ and counter-ions $Na^{+}$.  The behavior of polyelectrolytes is
much less understood systems compared to others commonly discussed in
macromolecular science. The main reason is the difficulty in applying
renormalization group  and scaling ideas in systems where long ranged
forces are present. Moreover, the special features of the Coulomb potential
below (long ranged) and above four (short ranged) dimensions prevent from a
clear physical picture even in three dimensions.  For weakly charged and
flexible polyelectrolytes the situation is less severe
\cite{borue,jl,borsali}. First, the flexibility allows to use the
Wiener measure, i.e., Gaussian chains in absence of interactions, and
electrostatic corrections to the stiffness of the chain \cite{odijk} can be
safely ignored. Second, in three dimensions the Coulomb potential is long
ranged ($1/r$) and variational techniques are able to provide physically
sensible results.

Variational methods have been discussed in the context of polymer physics
\cite{orland,cl,descloizeauxbook,us,bouchaud} and these statements have been
carefully investigated, with the result that variational methods are well
suited to study systems with long ranged interactions.  In the last two
decades considerable effort have been devoted to the statistical properties of
polyelectrolyte solutions, \cite{Oosawa,Gennes,Pfeuty}. It is usual custom to
study strongly charged macromolecules, where the Coulomb interaction between
the monomers is larger than the short range interactions of the van-der-Waals
type.  Here we will investigate both regimes by means of an extended
variational method, including the case of weakly charged polyelectrolyte
chains \cite{khokhlov,khokhlov2}.  We are going to study the collapse of a
single chain, when the short ranged interactions dominate and recover the
limit of strongly charged chains, where elongated conformations are expected
\cite{khokhlov2,joanny}.

The main challenge is to study the case of poor solvent conditions and the
formation of globular states. So far, these regimes are mainly studied by
scaling considerations and blob pictures \cite{brochard,johner,vilgis} as
well as Monte Carlo simulations \cite{kremer} and by mean field approaches
\cite{Grosberg,Grosberg2}.  The essential features of a neutral chain in poor
solvent can be summarized as follows. Attractive and repulsive interactions
(two and three body terms in the virial expansion) balance each other; as a
result a globule of size $R \simeq b (N/\tau)^{1/3}$ forms, where $\tau$
defines the relative temperature distance from the theta point, $\tau =
|T-\Theta|/\Theta$, and determines the second virial coefficient $v = -
b^3\tau$. The globule itself can be viewed as a melt of blobs of diameter $\xi
= b/ \tau$. The density of the globule is then given by $c = \tau/b^3$. The
main difficulty, however, is to obtain these properties from a more rigorous
point of view. This seems to be nontrivial, since most of the properties of
the globule are determined by the surface tension $\gamma_0 \simeq k_{\rm
  B}T/\xi^2$.  However, variational techniques are suitable for these cases,
since fluctuations inside globules are small, and play only a significant role
on length scales smaller than the thermal blob size $\xi$, where entropy
dominate and the chain conformation is Gaussian. The last remark applies also
for the surface of the globule.

In the simple case of neutral chains, we investigate the coil to globule
transition driven by the short ranged interactions.  Globule formation, as a
balance between the two and three body terms, takes place under poor solvent
conditions.  The physics of uncharged macromolecules has been nowadays well
understood \cite{Gennes,descloizeauxbook,Grosberg,Grosberg2}.  Charged
macromolecules are far less understood systems because of the interplay of
short ranged attractive forces and long ranged Coulomb forces.  It is our
first aim to recover the scaling results for a neutral chain within the two
variational schemes introduced in section II.  Previous attempts to describe
poor solvent chains via a variational method \cite{sodeberg} and the
description of the necklace phase \cite{olvera} should be mentioned as well.
Once the correct scaling behavior is recovered, we will return to the problem
of a weak Coulombic perturbation and its effect on the macromolecule
conformations. This is discussed in detail in section III and IV.  As expected
from standard energy considerations \cite{rubinstein,kardar1,kardar2} the
globule is stable for charges smaller than a critical threshold (in a similar
fashion as a charged liquid drop). For charges exceeding this point, the
globule will elongate to minimize the Coulomb repulsion. Due to connectivity
of the macromolecule, the globule will split into two which are connected by a
stretched string of thermal blobs.  In the following we will explore this
scenario within the variational method for an open chain in section IV.  The
properties of the necklace phase are discussed and a global phase diagram in
the space of solvent quality and inverse Bjerrum length is presented.

\section{Polyelectrolyte Phenomenology}

\subsection{Polyelectrolyte Chain in Theta Solvent}

In this section we review the fundamental scaling results for
polyelectrolyte chains in the case of a dilute solution below the
overlap concentration, such that only single chain statistics can be 
taken into account \cite{joanny,Gennes2}.
We assume first to be in the weakly charged limit, where the
charge fraction is small enough to assume Gaussian statistics for the non
interacting chain and to ignore counter-ion condensation.  At high
temperature, the interaction between the counterions and polyelectrolyte
chains in solution is weak and the distribution of counter-ions is
homogeneous. Only a negligible small fraction of counter-ions are in the
neighborhood of the chain, and their influence on the chain conformation can
be neglected. Chain entropy and Coulomb interaction will then
determine a globally stretched  conformation of the chain. At smaller
scales, entropy dominates the behavior and the usual blob
picture holds \cite{Gennes2}. The chain is made of a string of electrostatic
blobs of size $ \xi_{\rm el}$, each made of $g_{\rm el}$ monomers.  The size
of each blob and the number of monomers within one blob can be determined by
simple energy considerations.  The electrostatic energy of a
blob is of the order of the thermal energy, i.e. $(fg_{\rm el})^2 l_{\rm B}/
\xi_{\rm el} \simeq 1$ and the Bjerrum length is defined as $l_{\rm B}= e^2 /
\epsilon k_{\rm B}T$, where $ \epsilon$ is the dielectric constant of the
solvent and $T$ the temperature.  Moreover the polymer segments inside the
blob obey Gaussian statistics, i.e.  $ \xi_{\rm el} = b g_{\rm el}^{1/2}$, so
that the size of the blob is given by $\xi_{\rm el} \simeq b (b/l_{\rm
  B})^{1/3}f^{-2/3}$. The length of the polyelectrolyte is then $R \simeq
(N/g_{\rm el}) \xi_{\rm el}$, so that
\begin{equation}
R \simeq b (l_{\rm B} f^2/b)^{1/3}  N
\label{length}
\end{equation}
A numerical solution of the variational method 
described in section II, confirms the scaling behavior predicted by equation
(\ref{length}), including logarithmic corrections in the
form $N(\log N )^{\gamma}$ Moreover, we were able to derive equation
(\ref{length}) \cite{us}, by assuming the proper ansatz for the
monomer--monomer correlation function at large distances, assuming a non
universal dependence of the exponent $\gamma$ at small values of the
interaction strength. Differently, a mere implementation of des Cloizeaux
asymptotic analysis for polyelectrolytes would lead to an unphysical
overstretching, i.e., 
$R\propto N^2$ for the end to end distance.
Details are described elsewhere \cite{us,allegra}.
Presence of logarithmic corrections to equation (\ref{length}) were
already suggested in the past \cite{Gennes2,allegra,kachalski}, with a
universal exponent $ \gamma = 1/3$.
The presence of logarithmic corrections, that makes the blob size a
non-universal quantity, appears naturally in the variational
approach and finds its origin in the mean field nature of the method, 
similarly to logarithmic corrections in a field theory at the upper and lower
critical dimension. The electrostatic blobs close to the
end of the chain become larger, since end monomers experience less
electrostatic interactions compared to monomers in the middle of the
chain, leading to inhomogeneous fluctuations in the electrostatic potential.
These effects have been recently studied by an analysis of the
classical path of the polyelectrolyte chain \cite{joanny2}.
The resulting generalized expression for the blob size reads
\begin{equation}
\xi_{z} = \frac{ \xi_{\rm el}}
{ [ \log ( \frac{ (L/2)^2 - z^2 }{ L \xi_{\rm M} } ) +1 ]^{1/3} }\quad,
\label{xiz}
\end{equation}
where $\xi_{\rm el}$ is the usual electrostatic blob size, obtained
from scaling arguments. The quantities $L$ and $\xi_{\rm M}$ are
respectively the length of the chain and the maximum blob size at the
chain ends. It is important to remark that, neglecting the $z$
dependence in equation (\ref{xiz}) one recovers the variational result
of de Gennes $et ~al.$ \cite{Gennes}, with the proper logarithmic
behavior. Shortly below, we show how these correlations
can be detected as well, by a more subtle variational technique, which uses
heterogeneous monomer--monomer correlation functions as trial
function for the chain statistics.

\subsection{Polyelectrolyte Chain in Poor solvent}

We consider first a polyelectrolyte chain in poor solvent, where the monomer
statistics becomes much richer, compared to that related to the theta solvent.
The polyelectrolyte chains consists of $N$ monomers of size $b$ and the
fraction of charged monomers $f$.  The uncharged chain in a poor solvent forms
a globule.  The monomer density $\rho = \tau b^{-3}$ inside the globule is
defined by the balance between the two body attraction and the three body
repulsion.  From standard scaling arguments we can distinguish four different
regions. At weak enough values of the solvent quality the chain is Gaussian
and $ R = b N^{1/2}$. Below the critical value of $\tau_c = N^{-1/2}$ we enter
the globular phase where the overall size of the chain scales as
\begin{equation}
R=b\left(\frac{N}{\tau}\right)^{1/3} \quad.
\label{glob}
\end{equation}
We review briefly the scaling theory when Coulomb interactions  are present
to the poor solvent chain
\cite{khokhlov,rubinstein,schiessel,pincus}. For
values of the two body attraction lower than the critical threshold $\tau <
\tau_c$ the chain will extend to the electrostatic blob chain discussed above
if the electrostatic interaction exceeds $l_{\rm B}f^{2}/b > N^{-3/2}$, and
therefore overcomes the Gaussian entropy.  In the poor solvent regime $\tau >
\tau_c$ the globules remain spherical for small charge fractions. However, if
the charge fraction increases and exceeds the value
\begin{equation}
u^{*}\equiv\frac{l_{\rm B}^{*}f^{2}}{b} > \frac{\tau}{N}
\label{inst}
\end{equation}
the globule changes shape and an elongated structure emerges, still
maintaining the local scaling properties of a dense globule.  The extended
globule is referred in the literature as a cylindrical globule
\cite{rubinstein}.  In absence of charges the surface tension $
\gamma_0$ of the globule is of the order of $kT$ per blob at the
globular surface, i.e.  $\gamma_0\approx k T \xi^{-2}$.  
When the charge fraction becomes larger, the Coulomb
repulsion between the monomers will finally dominate over the surface free
energy therefore affecting the shape of the globule.  Whenever the Coulomb
repulsion $F_{\rm coul} \approx k_{\rm B} T l_{\rm B}f^2N^2/R$ becomes
comparable to the surface energy $F_{\rm sur} \approx   \gamma_0 R^2
\propto k_{\rm B} T N^{2/3} \tau^{4/3} $, the
critical Bjerrum length at which this occur can be simply derived by comparing
the bulk and surface free energy of the globule given above \cite{kardar2}.
In this regime the average length of the cylindrical globule is given by

\begin{equation}
R_{\parallel} = b {N \over \tau} \left( \frac{l_{\rm B} f^2}{b}\right)^{2/3}\quad.
\end{equation}

At higher values of the critical Bjerrum length defined by eq.(\ref{inst})
the globule eventually split into two and more connected
globules, according to a Rayleigh instability of charged spheres
\cite{kardar1,kardar2,rubinstein}.
At even larger charge fractions the chain finally stretches and the
standard electrostatic blob regime is recovered, as soon as
\begin{equation}
u^{**}=\frac{l_{\rm B}^{**} f^{2}}{b} > \tau^{3},
\label{inst2}
\end{equation}
the chain becomes extended and the electrostatic and thermal blob size become
equal. The chain has then an extension $R=bN\tau$. The foregoing scenario is
depicted in Fig.1 for a ring and an open chain, neglecting for the moment a
necklace formation, as an alternative, energetically favorable conformation to
the cylindrical globule introduced by Khokhlov.  Below, we investigate these
different regimes predicted by scaling theory using a new variational
technique we recently introduced \cite{us}.  It is important to remark that
connectivity plays here a crucial rule.  All sort of scaling arguments do not
take connectivity into account, while we will be able to describe the globule
to necklace transition by mean of a variational method that starts from the
original Hamiltonian for the problem and properly include the connectivity of
the chain.  We consider a generalized form of the Gaussian variational method
\cite{cl,descloizeauxbook} to study the scaling properties and spatial
conformation of a polyelectrolyte globule in poor solvent, for different
values of the Coulomb interaction strength between the monomers, obtaining an
independent test for the results commonly discussed in the scaling theory
approach.  By mean of the generalized Gaussian variational method we will be
able to study the overall extension of the globular structure measuring, e.g.
the end to end distance and the internal properties of the extended globule,
e.g. point to point correlation function for the monomer positions. The
essential physics in terms of the variational problem will be described by the
Euler equations corresponding to free energy minimization. Although important
conclusions concerning the overall size of the chain can be obtained \cite{us}
via the asymptotic analysis of the Euler equation, we will mainly pursue in
this paper a numerical solution for finite number of monomers $N$.

In section II we review the variational method as it was discussed previously
for homopolymer chains and use a more generalized form to polyelectrolyte
chain in theta and poor solvent \cite{cl,descloizeauxbook,us}.  We will show
how the different variational approach can be successfully applied to either
open chains or chains with a cyclic geometry.  The cyclic invariance \cite{cl}
has the advantage to reduce the computational effort required to minimize our
variational equations, because the chain coordinates can be decomposed in
orthogonal (Rouse) modes. For open chains, the end effects have to be included
explicitly.  In section III present the results for cyclic chains of increasing
size $2^4<N<2^{16}$. A global phase diagram in terms of solvent quality and
inverse Bjerrum length is given.  In section IV we consider the variational
approach for open chains and we present the results for chains of length
$2^4<N<2^9$.  The open chain case result to be very involving, since the
computational effort is enhanced of a factor $N/ \log_2 N$ with respect to the
cyclic chain case, but inhomogeneous point to point correlation function are
obtained and discussed. These will reveal the end effects we discussed for
polyelectrolyte chains in theta solvent. Moreover, in the extended globule
region, a non-monotonic correlation function, showing structural modulation
along the chain was found, reminiscence of necklace formation.

\begin{figure}
  \begin{center}
    \includegraphics[scale=0.35,angle=0]{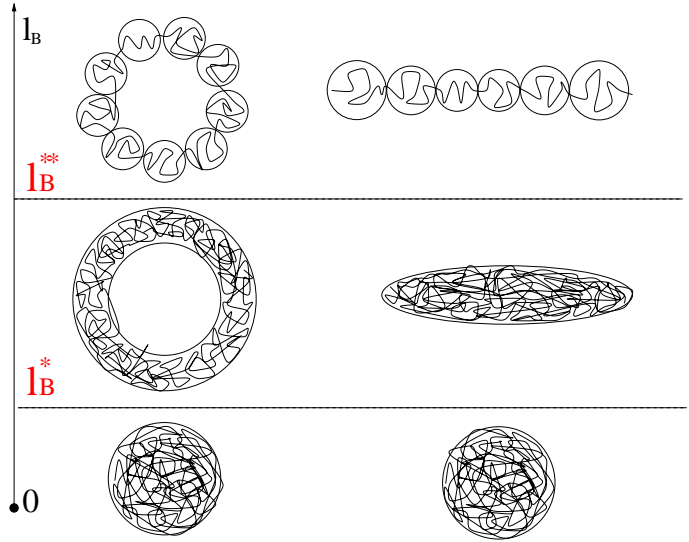} \vspace{10pt}
  \caption{For the two architectures considered, i.e., open chains
    and ring chains, we expect for increasing values of the interaction
    strength, $f^2 l_{\rm B}$ a first transition (see equation (\ref{inst})
    from globule to extended globule and a second one, at higher values of the
    interaction strength from extended globule to open chain (see equation
    (\ref{inst2}).}
   \end{center}
\end{figure}
\section{The Gaussian Variational method}
\subsection{The homogeneous variational method - cyclic chains}
In this section,  we apply the variational method (Gaussian
approximation) to study  the conformation of a polyelectrolyte chain 
in poor solvent. We compute the free energy of a self-interacting
polymer chain with Coulomb repulsion between the monomers, based on a 
discrete representation of a chain of $N$ monomers. The method relies 
on the well known variational principle, where a Gaussian trial
probability is proportional to the exponential of a quadratic form of 
the monomer coordinates $\vec{r_j}$.
The variational free energy for a uniformly charged polymer of length $N$ in a
$d$ dimensional space can be computed according to the usual Gibbs
Bogoliubov inequality
\begin{equation}
F \le F_V = \langle H - H_0 \rangle_{0} +F_0,
\label{Inequal}
\end{equation}
where $F_0$ is the free energy of the Gaussian model defined by equation
(\ref{Gauss}).  In equation (\ref{Inequal}), $H$ represents the full
Hamiltonian of a charged polymer chain with a long - range
monomer--monomer interaction,
\begin{equation}
 V(\vec{r_i} - \vec{r_j}) = | \vec{r_i} - \vec{r_j}| ^{ -\lambda},
\label{potential}
\end{equation}
and where is generally $\lambda < d$. However throughout the present
paper we use only the three dimensional Coulomb case, i.e.,
$\lambda=d-2, d=3$, see also Appendix A.
Following the Cloizeaux' \cite{cl} the expression for the Hamiltonian reads
\begin{eqnarray}
H/k_BT = \frac{d}{2b^2} \sum_{i=1}^N ( \vec{r}_{i+1} - \vec{r}_i )^2 - |v|/2
\sum_{i=1}^N \sum_{j=1}^N  \delta(\vec{r}_i - \vec{r}_j )+ \nonumber \\
w/3! \sum_{i=1}^N \sum_{j=1}^N\sum_{k=1}^N  \delta(\vec{r}_i - \vec{r}_j )
   \delta(\vec{r}_j - \vec{r}_k )+
l_Bf^2/2 \sum_{i=1}^N \sum_{j=1}^N V(\vec{r}_i - \vec{r}_j),
\end{eqnarray}

For the variational free energy $F_V$ in the limit of long
chain $N \gg 1 $,

\begin{eqnarray}
\frac{F_V[g(k)]}{k_BT N} &=&\frac{d}{2} b(1)   
-\tau \sum_{n=1}^{N}
[b(n)]^{- \frac{3}{2}} 
+ w\left(\frac{d}{2\pi}\right)^{3}\sum_{n=1}^{N}\sum_{m=1}^{N-n}
[b(n)]^{- \frac{3}{2}}[b(m)]^{- \frac{3}{2}} \nonumber \\
&+& u C_h(3,1)\sum_{n=1}^{N}[b(n)]^{-\frac{1}{2}} +\frac{d}{4 \pi} \int_{ - \pi}^{ + \pi} \log g(k) dk - \frac{d}{2} [ 1+ \log (2 \pi/d) ]
\label{Euler4}
\end{eqnarray}
where   
$b(n)\equiv {1}/{b^2} ~ \langle ( \vec{r_i} -\vec{r_{i+n}})^2 \rangle $ 
and is related to $g(k)$  via equation (\ref{A})
and $C_h(3,1)$ is a constant depending on the dimension $d$  and
interaction exponent $\lambda$. 
We  also  express  the coupling constants as  dimensionless
quantities $u,\tau,w$. 
The electrostatic interaction strength $u$ is defined in term of 
Bjerrum length and charge fraction $ u  \equiv l_{\rm B} f^2/b$.
The second virial coefficient is replaced by $\tau = v/b^3$.   
The third virial coefficient  $w$ is fixed to $b^6$.
During all numerical evaluations set  $w/b^6 \equiv 1$ for simplicity and
normalization.     
Minimization of the variational free energy $F_V$ yields
\begin{eqnarray}
g(k) &=& (1- \cos(k)) \nonumber \\ 
&-& \tau \left(\frac{3}{2\pi}\right)^{\frac32}\sum_{n=1}^{ N } (1 -
\cos(nk)) [b(n)]^{- \frac{5}{2}} \nonumber \\
&+& w \left(\frac{3}{2\pi}\right)^{3}\sum_{n=1}^{ N} \sum_{m=1}^{N-n}
\{ (1 - \cos(nk)) [b(n)]^{- \frac{5}{2}} [b(m)]^{- \frac{3}{2}}
+ (1 - \cos(mk)) [b(m)]^{- \frac{5}{2}} [b(n)]^{- \frac{3}{2}} \}
\nonumber \\
&+& u \frac{ C_h(3,1)}{3}
 \sum_{n=1}^{ N } (1 - \cos(nk))[b(n)]^{-\frac{3}{2}}.
\label{Euler2tris}
\end{eqnarray}
Equation (\ref{Euler2tris}) is a self consistent equation for $g(k)$.
The variational principle we discussed is characterized by isotropic
correlation functions, as we assumed (see Appendix A) considering
Fourier representation. This is of course related to the cyclic chain
architecture we assumed throughout the calculation.
In the extended globular phase, onset of the Rayleigh instability,
and the formation of necklaces, these isotropic and homogeneous
correlation functions appear no longer sufficient to describe a
structure, that, if shown to exist, is intrinsically inhomogeneous.
Therefore we are going to generalize the above variational method for
heterogeneous structures, releasing the translational invariance
constraint.

\subsection{Inhomogeneous variational method for an open chain}

In this section we generalize the variational method to study an open chain 
(without assuming cyclic invariance) and in $d$ - dimensions. 
The free energy $F_N$ depends on the entire spectrum
of correlation function $Q(i,j) = \langle \vec{r_i}\vec{r_j} \rangle$.
Using the Gaussian trial Hamiltonian
$H_o = \sum_i\sum_j Q^{-1}(i,j)\vec{r_i}\vec{r_j}$ we derive $F_N$ in theta solvent,    
\begin{eqnarray}
F_N/k_BT&=& -\frac{d}{2}{\rm Tr} \log \hat{Q} - \frac{d}{2} N +
\frac{d^2}{2b^2} \sum_{n=1}^N [Q(n,n)+Q(n+1,n+1)-2Q(n,n+1)] \nonumber \\
&+& u C(d)\sum\limits_{1\le m }\sum\limits_{< n \le N}
[Q(n,n)+ Q(m,m)-2Q(n,m)]^{- \frac{1}{2}},
\label{via}
\end{eqnarray}
where $C(d)$ is a constant which depends only on dimensionality (see Appendix).
In $d=3$ for Coulomb interaction ($\lambda=1$), $C(d) = (\Omega_3)^2
 (\sqrt{\pi/2})/(2\pi)^3$ and $\Omega_3 = 4\pi$ is the solid angle in $d=3$.  
The notation  $\hat{Q}$  stands for matrix. 
Minimization of  the free energy yields
\begin{equation}
\frac{\delta F_N}{ \delta Q(i,j) } = 0
\end{equation}
and  we obtain the Euler equation for an open chain  corresponding 
to equation (\ref{via}) in the form of a $N\times N$ matrix equation.
\begin{eqnarray}
\frac{d}{2} \hat{Q}^{-1}(i,j) &=& \frac{d^2}{2b^2} P(i,j) - \frac{u}{2} C(d)
\sum_{n\neq i}[ Q(i,i)+ Q(n,n) -2Q(i,n) ]^{ -\frac{3}{2}}
\delta_{ij} \nonumber \\ \label{inh}
&+& \frac{u}{2} C(d)[ Q(i,i)+Q(j,j) -2Q(i,j) ]^{ -\frac{3}{2}}
(1-\delta_{ij}),\\
P(i,j) &=&  cn(i)\delta_{ij} - \delta_{i,j+1} - \delta_{j,i+1}.
\end{eqnarray}
where  matrix  $P(i,j)$  expresses connectivity in terms  
of the monomer coordinates $\vec{r}_i$.
The ends of the chain  satisfy different connectivity condition, $cn(i) =1$ 
for monomers at the end of the chain $i=1,N$ and $cn(i) =2$ for monomers
along the chain.  
A general expression of the free energy for arbitrary solvent conditions 
is presented in Appendix B. 
The right hand side of Eq.\ref{inh} defines  each element of the matrix 
$\hat{Q}^{-1}$. This matrix equation has been solved
self-consistently.  The  computational cost of matrix inversion 
scales like $N^3$.
Note that the sum of each column (row) of $\hat{Q}^{-1}$ is vanishing
(since the diagonal element $Q_{ii} = -\sum_{j\neq i} Q_{ij}$)
indicating the linear dependence of the columns(rows).  Due to the 
translational invariance the eigenvalue of matrix has a zero mode 
corresponding to the motion of the center of mass. 
This cause the determinant of the matrix to vanish.
It is indeed necessary to introduce the constraint 
$\vec{r}_N = -\sum_i \vec{r_i}$ to 
eliminate the degree of freedom corresponding to 
the motion of the  center of mass.
Proceeding to the first iteration of the self consistent Eq.\ref{inh}, 
we define the reduced connectivity $(N-1)\times(N-1)$  matrix  $\hat{N_r}$  
in the center of mass coordinate system.
This can be obtained by expressing $\vec{r}_N$ in terms of ${\vec{r}}_i$'s
(i=1,..N-1), according to the following identity.
\begin{equation}
\sum_i^{N}\sum_j^{N} P(i,j)\vec{r}_i \vec{r}_j =
 \sum_i^{N-1}\sum_j^{N-1}
P_r(i,j)\tilde{r}_i\tilde{r}_j  
\end{equation}  
We obtain 
\begin{equation}
P_r(i,j)=P(i,j)+P(N,N)-P(i,N)-P(N,j), \,\,(i=1,..N-1).
\end{equation}
This operation still preserves the symmetry $P_r(i,j) = P_r(j,i)$.
We include the reduced form of the connectivity matrix $P_r$ (see 
Eq.\ref{inh}) and start the iterative scheme discussed above.    
Another way to remove the zero mode is by using the relative
coordinate system (bond vector coordinates)
$\vec{b}_i = \vec{r}_i -{\vec{r}_{i-1}}$ which is the case described in 
\cite{jonsson}.

\section{Results - Cyclic Chain}
In this section we present the results by solving  
the Euler equations
(\ref{Euler2}) and (\ref{Euler4}) for a polymer chain of length $N$ with
cyclic boundary conditions.
We  compute the optimal profile $g_{\rm opt}(k)$ that
minimizes the free energy $F_V$ (see eq. (\ref{Euler4})), for
increasing values of the interaction strength $u$, in a theta solvent.
At a given value of $u$ we can obtain a direct measure
of the blob size in the following way:
we compute the real space correlation function $b(n)$,  related via
Fourier transformation to the optimal profile $g_{\rm opt}(k)$ (see
 equation (\ref{A})).
In the weakly charged limit (i.e.
for small  values of $u$) the function $b(n) = n^{2\nu}$ shows a crossover from
Gaussian behavior (at small values of $n$, $\nu=1/2$) and a linear
behavior (at larger values of $n$, $\nu =1$)
as described in equation (\ref{length}).
We define the blob size as the extrapolated crossover point of the
exponent $\nu$  between
these two regimes.  The result is shown in Fig.2.

\begin{figure}
  \begin{center}
    \includegraphics[scale=0.45,angle=0]{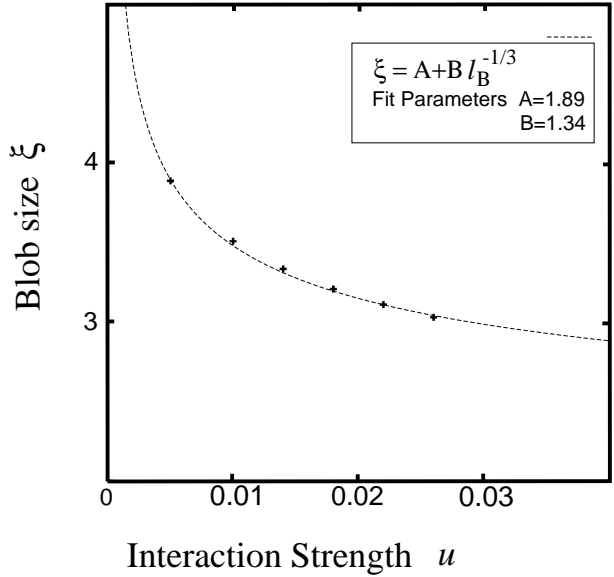} \vspace{10pt}
  \caption{Blob size, as measured from the direct solution of
   equations (\ref{Euler2}) and (\ref{Euler4}), considering the crossover
   between the small distance behavior of the mean square distance
   $b(n)$ versus the large distance behavior.  Different values of the
   electrostatic blob size for increasing values of the interaction
   strength $u$ are compared with the $\xi \approx l_{\rm B}^{-1/3}$
   behavior expected from phenomenological arguments.
   }
   \end{center}
\end{figure}

\subsection{Neutral chain in Poor solvent.}

We now investigate the poor solvent regime, first considering a
neutral chain. We will consider the effect of a Coulomb repulsion
between the monomers in the next paragraph. In poor solvent
conditions, as mentioned above, we expect the neutral chain to form a
globule. There is one important length scale in the globule, namely
the size of the density fluctuations. On length scales
larger than the correlation length $\xi_{\tau} \approx b g_{\tau}^{1/2}$, the
interaction between the monomers overcomes the thermal fluctuations,
resulting in a dense, liquid like structure, made of Gaussian blobs.
We were able to confirm $quantitatively$ this scenario within our
variational principle.
We start from a Gaussian chain and let the solvent quality decrease.
The density of the globule is determined by interplay of 2-body
attraction  and 3-body repulsion.   As in the case of a theta solvent,
 the monomer--monomer
correlation function shows a crossover between a Gaussian regime
$b(n) \propto n$ and
a regime where $b(n) \propto n^{2/3}$. Moreover, to check that our
variational method properly describes a neutral chain in poor solvent
conditions we measure the overall size of the globule as a function
of the solvent quality. One should expect to recover results that are
consistent with equation (\ref{glob}).

In Fig.3 the end to end distance is shown as a
function of $\tau$. As expected the globule is formed
as  solvent quality becomes poorer ($\tau$ increases) meanwhile the
three body term is constant, as usually assumed in scaling theory.

\begin{figure}
\begin{center}
\includegraphics[width=8cm]{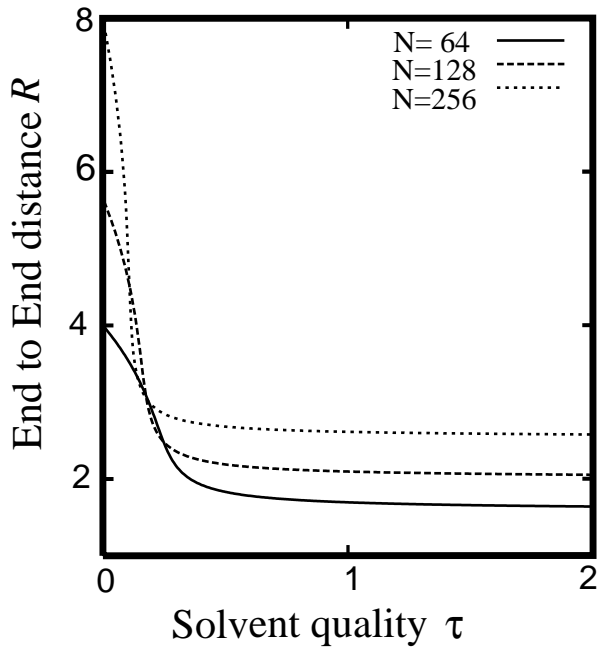}\vspace{10pt}
  \caption{End to end distance behavior in poor solvent for a cyclic 
  chain of length $N$. The end to end distance is obtained for 
different $\tau$ for chain length N=64,128,256. The entire spectrum 
of correlation functions $b(n)$ is measured at a given value of 
the solvent quality $v \simeq 0.4$. }
\end{center}
\end{figure}

In Fig.4 the globule size is shown. The proper scaling behavior
as a function of the chain length, predicted by scaling
(see equation (\ref{glob})), is recovered.

 \begin{figure}
  \begin{center}
    \includegraphics[width=8cm]{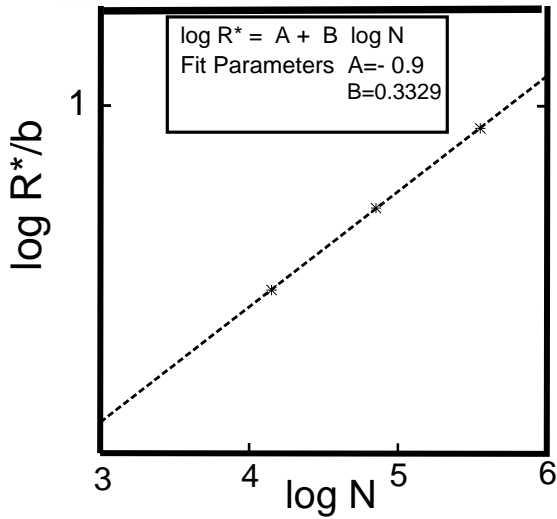} \vspace{10pt}
 \caption{Globular End to end distance R* versus chain length for
  chains of increasing length $N=64,128,256$.
  The continuous line is obtained by a fit of our numerical results,
 where the fit parameters are $A=-0.9$ and $B=0.3329 \pm
 0.0002$. Similar results can be obtained for globules of different
initial density. The proper scaling of $R$ with density has also
been checked carefully and confirms scaling results.}
  \end{center}
 \end{figure}

We observe the overall size of the system to scale as $R \propto N^{1/3}$.
The same scaling behavior is observed when a weak enough Coulomb
interaction is included within our variational approach.
A sharp transition to extended globule appears for values of the
Bjerrum length exceeding a critical value $l_{\rm B}^{*}$ as shown in Fig.5.
According to the scaling theory of Khokhlov a sharp transition
occurs due to Coulomb interactions between the monomers when the
surface free energy becomes of the order of the free energy
contribution due to excluded volume interactions. The variational
method takes into account the surface terms responsible for the
extended globule transition by a proper rescaling of the
monomer--monomer correlation functions.

Simple scaling arguments suggests that the critical value of the
Bjerrum length where the transition occurs scales as $1/N$ where $N$ is the
number of monomers in the chain. We show in Fig.5 the measured value
of the critical Bjerrum length value as a function of the chain length
$N$. Scaling properties of the end-to-end distance of
the chain have been also discussed for the long - ranged intra-molecular
interaction $1/r^{\lambda}$ at an arbitrary spatial dimensionality
$d$ \cite{bouchaud}. The validity of the variational
approach for short range interacting homopolymer chains must be questioned
\cite{bouchaud} for natural reasons. In poor solvent the problem of
logarithmic corrections that spuriously appear within
the Gaussian Variational Principle cancels between the two and three
body terms, so that a power law behavior for the
end to end distance is measured, i.e. $R \propto N^{1/3}$, as in Fig.4.

\subsection{Polyelectrolytes chains in poor solvent}

We now consider  a polyelectrolyte
in  a poor solvent.
When the polymer is charged the Coulomb repulsion between the
monomers changes the shape
of the globule but not its overall volume, defined only  by the
solvent quality. When the Coulomb repulsion $F_{\rm coul} \approx
k_{\rm B}Tl_{\rm B}f^2N^2/R$ is comparable to the surface energy
$F_{\rm surf} \approx \gamma_0 R^2$ the globule  either elongates
into a cylindrical shape or  splits into two
connected globules.
We  expect an instability in the radius of gyration at finite value
of $u$  once we include a Coulomb repulsion in our
variational equations.
Such a scenario is captured in the numerical implementation of the
Euler equations (\ref{Euler2}). For a given chain length, the
radius of gyration $R$ has a sharp increase at a critical  value of $u^*$.
According to Eq.\ref{inst} we expect $u^*$ to scale as
$1/N$.  This scaling behavior is shown in Fig.6.

\begin{figure}
  \begin{center}
    \includegraphics[width=8cm]{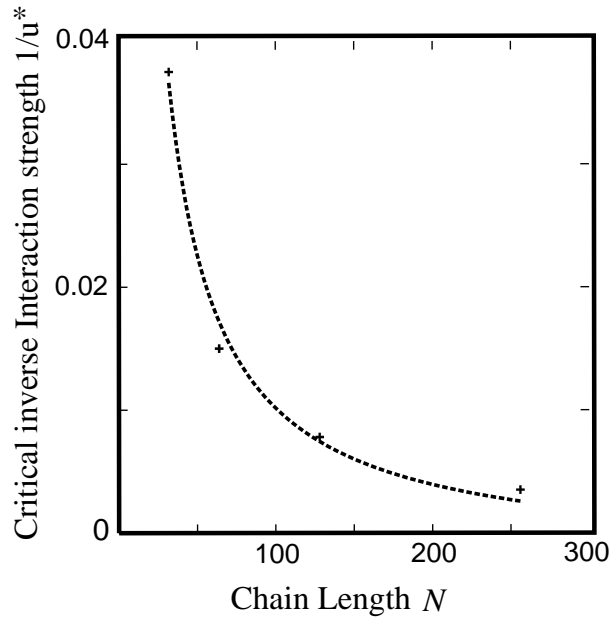}\vspace{10pt}
\caption{critical interaction strength $u^*$ versus chain length $N$.}
  \end{center}
\end{figure}

Fig.6  shows this instability for a given chain of lengths
$N=64$ and $N=128$. The end to end distance and its derivative are
computed as a function of $u$.

\begin{figure}
  \begin{center}
    \includegraphics[width=8cm]{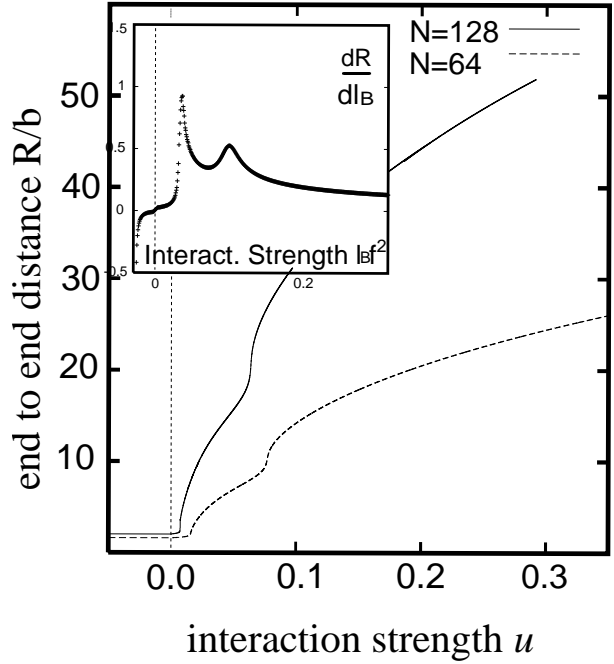}\vspace{10pt}
  \caption{End to End distance $R$ for increasing values of $u$.
  The inset on the top left corner shows
  the derivative of $R$ vs $u$.}
\label{phase_transition}
\end{center}
\end{figure}

Following the sharp transition point in Fig.6
allow us to construct the phase diagram for poor solvent conditions.
The phase diagram, for different values of the interaction 
strength $u$, is shown in Fig.7.  

\begin{figure}
  \begin{center}
    \includegraphics[width=8cm]{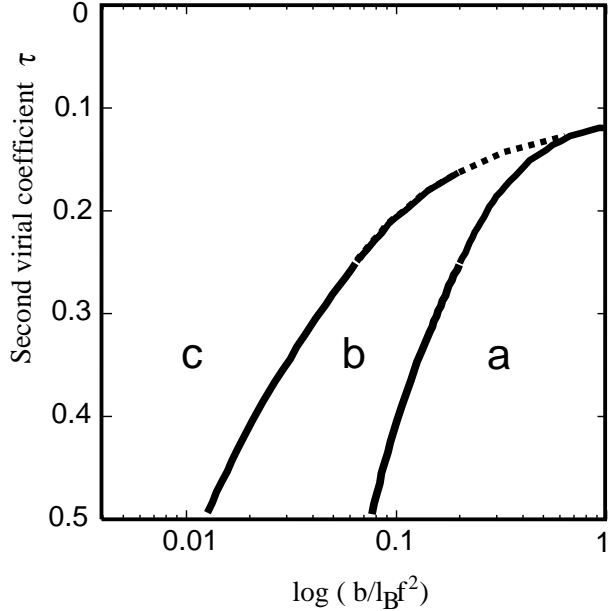} \vspace{10pt}
  \caption{Phase Diagram in poor solvent regime  and inverse
    Bjerrum length, for a chain of length $N=2^8$. phase boundaries
are represented by solid lines. The dashed line indicates that the
transition becomes very weak according  the hight of the second peak of Fig.
6. the two boundaries on the phase diagram
discriminate: (a) the neutral globule phase, where the polyelectrolyte
is under-charged and follows simple neutral chain statistics.
(b) the cylindrical globule phase, where a multiple crossover for the
monomer--monomer correlation function is observed from Gaussian
($b(n) \propto n , n< \xi_{\tau} $), to dense ( ($b(n) \propto
n^{2/3}, N >> n > \xi_{\tau}$) to extended ($b(n) \propto n^2, n \le N $).
(c) the stretched chain regime, where electrostatic blobs forms and
the chain is fully stretched }
  \end{center}
\label{phase_dia}
\end{figure}

In Fig.7 we recover the different regimes described in the
discussion of section I. The two boundaries on the phase diagram
discriminate: (a) the neutral globule phase, where the polyelectrolyte
is undercharged and follows simple neutral chain statistics.
(b) the cylindrical global phase, where a multiple crossover for the
monomer--monomer correlation function is observed from Gaussian
($b(n) \propto n, n< \xi_{\tau} $), to dense ( ($b(n) \propto
n^{2/3}, N >> n > \xi_{\tau}$) to extended ($b(n) \propto n^2, n \le N $).
(c) the stretched chain regime, where electrostatic blobs forms and
the chain is fully stretched (see Fig.2).

\section{The inhomogeneous variational method}

\subsection{Theta Solvent}

The application of Gaussian variational method to an open chain 
geometry allows us to investigate the end effect of finite chain
which was neglected the cyclic geometry assumption of section III. 
In a theta solvent,  the weakly charged polyelectrolyte 
has non uniform  blob size distribution\cite{joanny2}
indicating  charge depletion at the center of the chain.
In Fig.8 we compute the size of the blob at different relative 
position of the chain. 
The estimation of blob size is obtained finding a  
crossover point from Gaussian to extended chain statistics  
as explained earlier. The dashed line shows the  estimation 
from Eq.(1.2).

\begin{figure} logarithmic corrections in the
form $N(\log N )^{\gamma}$
  \begin{center}
    \includegraphics[width=8cm]{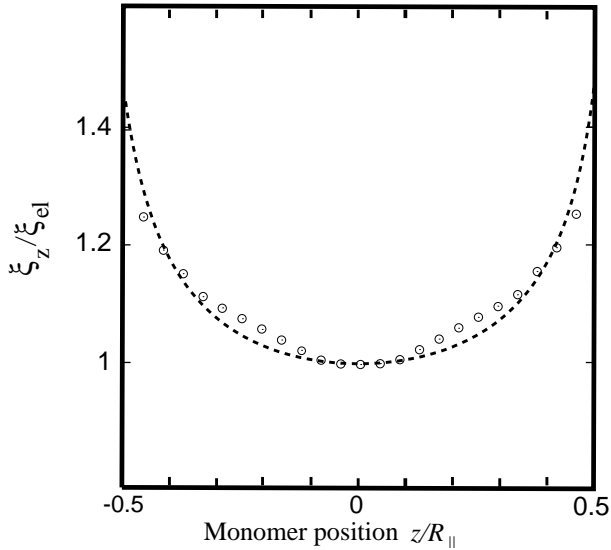} \vspace{10pt}
  \caption{ Blob size $\xi(z/R_{\parallel})$  at 
relative position $z/R_{\parallel}$.  The dashed line represents the
theoretical prediction from Eq.1.2 with $\xi_M/R_{\parallel} = 0.25$ }
\end{center}
\label{blobsize2}
\end{figure}

\subsection{Poor solvent. Necklace formation}

In the  cylindric globule phase, which we found
(within the variational method for cyclic geometry) of section IIIB,
we expect a correlation function that differs from the one computed in 
the globule phase.  
Any indication of globule splitting into sub-globules
connected by strings will appear as 
a modulation along the chain of the monomer to monomer correlation
function.The monomer--monomer 
correlation function $b(i,j) = Q(i,i)+Q(j,j)-2Q(i,j)$, in the globular
phase, is almost constant in the middle of chain, as is shown as a lowest curve in Fig.9.  
At the two ends of the chain, $b(i,n)$ increases and  
this is  an end effect in  a similar way as discussed in the 
case of a chain in theta solvent. At higher values of the interaction 
strength $u$, we observe (see Fig.9) two minima appears  
at both ends of the chain  and the maximum  
at the center part  of the chain.  The minimum (maximum) in the correlation
function corresponds to  the density maximum (minimum).    
It is clear that for large value of $u$, the center part of chain is 
more stretched out and connects to two  dense parts of the chain (pearls).  
The instability of the  globule  structure
manifests  clearly at the transition point $u^* \simeq 0.24$. 
The fluctuation of $b(i,n)$ is relatively large as can be seen by the curve 
in Fig.9.

\begin{figure}
  \begin{center}
    \includegraphics[width=10cm]{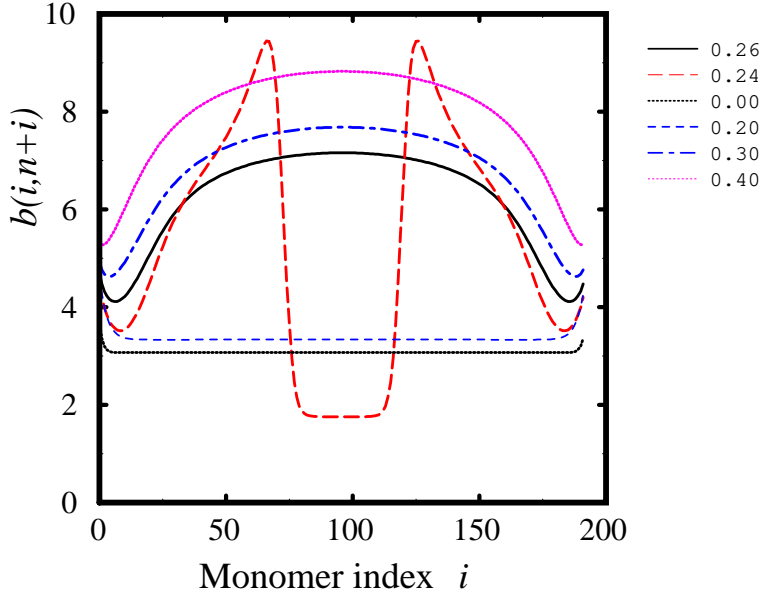} \vspace{10pt}
  \caption{
The correlation function $b(i,n+i)$, for various interaction strength $u$, is 
measured for fixed value of $n=7$.  The necklace formation proceeds 
the instability of the globular phase.   
  }
   \end{center}
\label{correlation}
\end{figure}

The  density maxima move to the edges of the chain
with increasing $u$. We interpret this as a structure
consisting  of smaller pearls connected by strings. 
To investigate the detail structure of this new scenario,  
we plot in Fig.10 $b(i,i+n)$ for different value of $n$ 
at the given interaction strength ($u = 0.26 > u^*$)
 for a chain of length $N=200$.  For intermediate values of $n$, 
 ($n\simeq 5$),  the density minimum clearly appears. 

\begin{figure}
  \begin{center}
    \includegraphics[width=10cm]{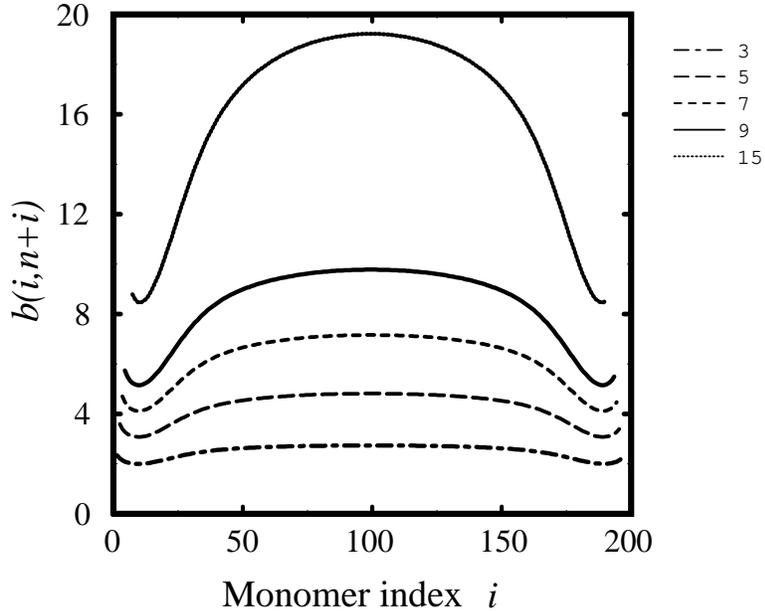} \vspace{10pt}
  \caption{Monomer-- monomer correlation function for values of 
        the interaction strength 
   $ u^{*} < u < u^{**}$
   for a chain of length $N=200$. The figure shows $b(i,n)$ vs $i$ 
   at different values of $n=3,5,7,9,15$}
\end{center}
\label{correlation2}
\end{figure}

\section{Conclusions}

We have shown in this paper that typical surface dominated physical effects in
polymer globules can be derived beyond scaling theory. Variational techniques
have been proven to provide satisfactory results for systems with long ranged
interaction potentials. Moreover, polymeric globules are systems where the
density fluctuations inside the globules are small and are ideal candidates
for variational computations. We have shown indeed a neutral globule can be
described within our variational scheme, as well as the repulsive instability
driven by Coulomb repulsion. The variational approach introduced in section
IIA, and usually encountered in the literature to discuss a self interacting
chain, suffers from the cyclic constrain that naturally hides end effects and
structural modulation along the chain. For this reason we generalized the
variational principle to the case of an open chain. The entire spectrum of
monomer-to-monomer correlation functions was indeed obtained in this way. As
mentioned above, the case of an open chain appeared to be more involved at the
computational level, so that the chain length considered was one order of
magnitude smaller compared to the case of a cyclic chain. Further improvement
of our numerical techniques would eventually reveal a multiple splitting of
the globule into sub-globules, beyond the simple ``dumbell'' formation we
observed.

\section*{Acknowledgments}

We would like to thank R. Everaers, J. F. Joanny, A. Johner, and H. Schiessel
for continuous and interesting discussions. N. L. acknowledges the grand by
the program ``Polyelektrolyte''of the German Science foundation DFG. T.A.V.
acknowledges support by the Laboratoire Europ\'een Associ\'e, L.E.A.  .

\begin{appendix}

\section{Variational approach under cyclic invariance condition}

Consider a Gaussian  variational probability distribution
\begin{equation}
P_V({\vec{r}_1},..,\vec{r}_N) = Z_V^{-1} \exp \{-H_0(\vec{r}_1,..,{\vec{r}_N}) \},
\end{equation}

where $H_0$ is given by

\begin{equation}
H_0(\vec{r_1},..,{\vec{r}}_N) = \frac{d}{2} \sum_{i=1}^{N} \sum_{j=1}^{N}
G(i-j) ({\vec{r}}_i - \vec{r}_j)^2
\label{Gauss}
\end{equation}

and $Z_V$ is a normalization constant determined by the condition

\begin{equation}
  \int P_V(\vec{r_1},..,\vec{r}_N)d^d \vec{r}_1\cdots d^d \vec{r}_N =1.
\end{equation}

 By the appropriate
choice of periodic boundary conditions for the chain, the quadratic form $H_0$
can be diagonalized by introducing cyclic coordinates (Rouse modes) for the
monomer positions

\begin{eqnarray}
\vec{ \rho}_q &=& N^{-\frac{1}{2}} \sum_{j=1}^N \exp [~ i 2 \pi j q/N ]
\vec{r}_j \nonumber \\
~\vec{r}_j &=& N^{-\frac{1}{2}} \sum_{q=1}^N \exp [- i 2 \pi j q/N ]
\vec{ \rho}_q.
\end{eqnarray}

Due to cyclic invariance the Cartesian components of $\vec{\rho}_q$ satisfies

\begin{equation}
\langle \rho^{(j)}_q \rho^{(j')}_{q'} \rangle = \frac{1}{d}
\delta_{jj'} \delta_{q q'} g^{-1}(2 \pi q/N),
\end{equation}

where the function $g_N^{-1}(2 \pi q/N)$ is positive and discontinuous
function of $k=2 \pi q/N$, and is related to the Gaussian propagator $G$ of
equation (\ref{Gauss}) via the following expression

\begin{equation}
g(2 \pi q/N) = \sum_{n =1}^{N-1} G(n) [1-\cos( 2 \pi qn/N)],
\label{translation}
\end{equation}

The variational free energy $F_V$ in the limit of infinitely long
chain $N \rightarrow \infty $,
\begin{eqnarray}
\frac{F_V[g(k)]}{k_BT N}&=&\frac{d}{2b^2} b(1) + F_h+ u C_h(d,\lambda)
\sum_{n=1}^{N}[b(n)]^{- \frac{\lambda}{2}} \nonumber \\
&+&\frac{d}{4 \pi} \int_{ - \pi}^{ + \pi} \log g(k) dk
 - \frac{d}{2} [ 1+ \log (2 \pi/d) ],
\label{Euler4tris}
\end{eqnarray}
where
\begin{eqnarray}
  \frac{F_h[g(k)]}{k_BT N}= - \tau \left(\frac{d}{2\pi}\right)^{\frac{d}{2}}
 \sum_{n=1}^{N} [b(n)]^{- \frac{d}{2}}
+ w \left(\frac{d}{2\pi}\right)^{d}\sum_{n=1}^{N}
\sum_{m=1}^{N-n}[b(n)]^{- \frac{d}{2}}[b(m)]^{- \frac{d}{2}},
\label{Euler4bis}
\end{eqnarray}
and 
$C_h(d,\lambda)=C(d,\lambda)\times d^{\frac{\lambda}{2}}$.
For arbitrary $\lambda$, 
\begin{eqnarray}
C(d,\lambda) = 2^{\frac{\lambda}{2}-1} \Gamma(d-\lambda -
  1)\Gamma(\frac{\lambda}{2})\sin\left[\frac{\pi(d-\lambda-1)}{2}\right]
  (\Omega_d)^2/(2\pi)^d
\end{eqnarray}
where $\Omega_d = 2\pi^{\frac{d}{2}}/\Gamma(\frac{d}{2})$ is the solid
angle for d-dimensions.  
For Coulomb interaction $\lambda =d-2$,$C(d)$  simplifies to 
$C(d) = \Omega_d^2/(2\pi)^2 \sqrt{\pi/2}$.

The function $g(k)$ is related to $b(n)$  by 
\begin{equation}
b(i-j)= \frac{1}{N} \int_{- \pi}^{ +\pi} \frac{(1-
  \cos((i-j)k))}{g(k)}dk,
\label{A}
\end{equation}
being,
\begin{equation}
b(i-j) \equiv \frac{1}{b^2}\langle (\vec{r}_i - \vec{r}_j )^2 \rangle.
\label{Square}
\end{equation}
The function $g(k)$ satisfies the symmetry requirements
\begin{equation}
g(k)=g(-k),~~g(k+2 \pi)=g(k).
\label{Euler3}
\end {equation}
Minimization of the variational free energy $F_V$ yields
\begin{eqnarray}
\label{Euler2}
g(k) &=& (1- \cos(k)) 
+ u \frac{\lambda}{d} C_h(d)\sum_{n=1}^{N} (1 -
\cos(nk)) [b(n)]^{-\frac{\lambda+2}{2}}\nonumber \\
&-&\tau \left(\frac{d}{2\pi}\right)^{\frac{d}{2}}\sum_{n=1}^{N} (1 - \cos(nk)) [b(n)]^{- \frac{d+2}{2}} \nonumber
\\
&+& w \left(\frac{d}{2\pi}\right)^{d} \sum_{n=1}^{N} \sum_{m=1}^{N-m} \{ (1 - \cos(nk))
[b(n)]^{- \frac{d+2}{2}} [b(m)]^{- \frac{d}{2}}
+(1 - \cos(mk))
[b(m)]^{- \frac{d+2}{2}} [b(n)]^{- \frac{d}{2}}
\end{eqnarray}
which reduces to equation (\ref{Euler2tris}) for $\lambda =1$ and $d=3$.

\section{The variational method for the inhomogeneous open chain}
In this appendix we write explicitly the Euler equations for the
inhomogeneous variational method, including terms
corresponding to two and three body interactions. In this case the
monomer-to-monomer correlation function $Q(i,j)$ depends from both its
arguments, not just from the distance $i - j$, as in the cyclic chain
case.
It is convenient to choose the trial Hamiltonian $H_O$ in the form
\begin{eqnarray}
H_0({\vec r}_1, \dots ,{\vec r}_N)   =
\frac{1}{2}\sum\limits_{j=1}^N\sum\limits_{l=1}^N Q^{-1}(j,l){\vec r}_j{\vec r}_l
\label{Ham}
\end{eqnarray}

The straightforward calculations leads to the following result for the
variational free energy:
\begin{eqnarray}
F_{V}(Q) &=& -\frac{d}{2}{\rm Tr}
\log \hat{Q} -\frac{d}{2}N
+ \frac{d^2}{2b^2}\sum_{n=1}^{N-1} b(n,n+1) + \frac{u}{2}
C(d)\sum\limits_{m }\sum\limits_{n\neq m}
\left[b(n,m)\right]^{-\frac{\lambda}{2}}\nonumber\\
&-& \frac{\tau}{2} \left(\frac{1}{2\pi}\right)^{\frac{d}{2}}
\sum\limits_{m}\sum\limits_{n\neq m}
\left[b(n,m)\right]^{-\frac{d}{2}}
+ \frac{w}{6}
\left(\frac{1}{2\pi}\right)^{d}\sum\limits_{k}\sum\limits_{n\neq k}\sum\limits_{m\neq k \neq n}
\left[b(k,n)\right]^{-\frac{d}{2}}\left[b(k,m)\right]^{-\frac{d}{2}},
\end{eqnarray}
where
\begin{eqnarray}
b(n,m) = Q(n,n) + Q(m,m) -2 Q(n,m) \quad ,
\label{B}
\end{eqnarray}
 As in the homogeneous
variational method, we proceed to compute the Euler equations,
obtained minimizing the above expression for the free energy in terms
of the propagator $Q(i,j)$.

\begin{eqnarray}
\frac{\delta}{\delta Q(i,j)} F_{V}\{Q(i,j)\} = 0 \quad .
\end{eqnarray}
Given that $\delta \left[{\rm Tr}\log\hat{Q}\right] /\delta\,Q(i,j)
 = {Q^{-1}}(i,j)$
one gets
\begin{eqnarray}
\frac{d}{2} Q^{-1}(i,j) &=&  \frac{d^2}{2b^2} [ cn(i)\delta_{ij}
-\delta_{i,j+1} - \delta_{j,i+1} ] \nonumber \\
&-&  \frac{\lambda}{2}l_{\rm B} C(\lambda,d) \sum\limits_{n \neq i}^N
[b(i,n) ]^{ -\frac{\lambda + 2}{2}} \delta_{ij}
 + \frac{ \lambda}{2} l_{\rm B} C [b(i,j)]^{ -\frac{\lambda+2}{2}}(1-\delta_{ij}) \nonumber \\
&+& \frac{\tau d}{2}
\left(\frac{1}{2\pi}\right)^{\frac{d}{2}}\sum\limits_{n\neq i}^N [b(i,n)]^{ -\frac{d+2}{2}}
\delta_{ij}-  \frac{\tau d}{2}\left(\frac{1}{2\pi}\right)^{\frac{d}{2}}
 [b(i,j)]^{ -\frac{d+2}{2}} (1-\delta_{ij}) \nonumber \\
&-& \frac{w d}{6} \left(\frac{1}{2\pi}\right)^{3}\sum\limits_{n\neq
  i}\sum\limits_{m\neq i \neq n}[b(i,n)]^{-\frac{d}{2}}[ b(i,m) ]^{ -\frac{d+2}{2}}\delta_{ij}
- \frac{wd}{6} \left(\frac{1}{2\pi}\right)^{d}\sum\limits_{m
  \neq i}\sum\limits_{n\neq m}[b(m,n)]^{ -\frac{d}{2}} [ b(m,i) ]^{ -\frac{d+2}{2}}\delta_{ij} \nonumber \\
&+& \frac{w d}{6}\left(\frac{1}{2\pi}\right)^{d}\sum\limits_{m=1}^N [b(i,j)]^{
  -\frac{d+2}{2}}\left\{[b(i,m))]^{-\frac{d}{2}}  +
    [b(j,m))]^{-\frac{d}{2}}\right\}(1 - \delta_{ij}).
\end{eqnarray}

\end{appendix}

\end{document}